\documentstyle[12pt]{article}
\newcommand{\be}{\begin{equation}}
\newcommand{\ee}{\end{equation}}
\def\bq{\begin{eqnarray}}
\def\eq{\end{eqnarray}}
\def\n{\nonumber}
\catcode `\@=11
\catcode `\@=12

\title{\bf\huge Global monopole as dual-vacuum solution in 
Kaluza-Klein spacetime}
\author{Naresh Dadhich\thanks{E-mail : nkd@iucaa.ernet.in} \\
{\sl Inter-University Centre for Astronomy \& Astrophysics,}\\
{\sl Post Bag 4, Ganeshkhind, Pune - 411 007, India.} \\
L.K. Patel \\
{\sl Department of Mathematics,}\\
{\sl Gujarat University, Ahmedabad 380 009, India.}\\
R. Tikekar \\
{\sl Department of Mathematics,} \\
{\sl Sardar Patel University, }\\
{\sl Vallabh Vidyanagar 388 120, India.}}
    \date{}
    
\begin{document}
    \maketitle
    
     \begin{abstract} 
                                  
  By application of the duality transformation, which implies interchange of 
active and passive electric parts of the Riemann curvature (equivalent to 
interchange of Ricci and Einstein tensors) it is shown 
that the global monopole solution in the Kaluza-Klein spacetime is dual to 
the corresponding vacuum solution. Further 
we also obtain solution dual to flat space which would in general 
describe a massive global monopole in 4-dimensional Euclidean space and would 
have massless limit analogus to the 4-dimensional dual-flat solution.
     
    \end{abstract}

    \n PACS numbers : 04.20,04.60,98.80Hw

    \newpage

 One of us has proposed a procedure [1-5] to construct spacetimes dual to 
the vacuum solutions in 4-dimensions. The procedure involves resolution 
of the Riemann curvature in analogy with the electromagnetics into 
electric and magnetic parts, defining a duality transformation 
between the two kinds of electric parts, and then applying it to the 
"effective" vacuum equation to get the dual-vacuum equation which would 
yield 
dual solution. By the effective equation, we would mean an equation 
that is less restrictive than the vacuum equation yet gives the same vacuum 
solution. The effective equation and its dual can be written in terms of 
energy density measured by a static observer and null and timelike  
convergence density [6]. The vacuum solution would follow from the 
condition of vanishing of energy and null convergence density while its 
dual from vanishing of timelike and null convergence density. Under the 
duality transformation energy density goes to timelike convergence 
density while null convergence density goes to itself. The effective 
equation would thus have a direct physical meaning. \\

 It turns out that the dual solution to the Schwarzschild 
black hole describes a black hole with global monopole charge [1,7]. That 
is under the duality transformation black hole acquires a global monopole 
charge. This procedure works for the Reissner-Nordstr${\ddot o}$m and the NUT 
solutions [5,8]. In the 2+1 gravity, it leads to new classes of black hole 
spacetimes [9].\\

 In this paper we wish to show that the same procedure would work for 
the  5-dimensional Kaluza-Klein spacetime, and the global monopole 
solution obtained by Banerjee etal [10] could be obtained as solution to 
the dual equation. In the Kaluza-Klein spacetime, 
analogue of the Schwarzschild solution was obtained by Gross and Perry 
[11] and Banerjee etal [10] have obtainend a solution of a global monopole 
which amounts to putting a global monopole charge on the 
vacuum solution. In this paper we wish to show that the global monopole 
solution [10] is dual to the vacuum solution [11].\\

 The Riemann curvature can be resolved similar to the elctromagnetic 
field into electric and magnetic parts relative to a unit timelike 
vector. The electromagnetic parts would be given by second rank 
3-tensors orthogonal to the resolving vector and they are defined as 
follows [1]:
    \be
    E_{ac} = R_{abcd} u^b u^d,  \tilde E_{ac} = *R*_{abcd} u^b u^d
    \ee
    \be
    H_{ac} = *R_{abcd} u^b u^d,  \tilde H_{ac} = R*_{abcd}u^bu^d  \\ 
    \ee
\noindent where a star denotes the hodge dual in 5-dimensions and 
$*R_{abcd}$ indicates the left dual (on first 2-form) and $R*_{abcd}$ the 
right dual (on second 2-form). We shall call $E_{ab}$ and $\tilde 
E_{ab}$ as the active and passive electric parts. The former is 
anchored to non-gravitational energy distribution while the latter to 
the gravitational field energy. By the 
duality transformation we shall mean interchange of active and passive 
electric parts with magnetic parts remaining undistrubed. The 
electrogravity duality transformation is thus defined by [1],
\be
E_{ab}\longleftrightarrow \tilde E_{ab}.
\ee
    
 \n The contraction of Riemann tensor is the Ricci tensor and that of 
its double dual (when all but one indices are properly contracted [12]) 
is the Einstein tensor. This means that the above duality transfromation will 
take Ricci to Einstein and vice-versa. Henceforth we shall directly work 
with the Ricci and the Einstein tensors rather than the electromagnetic 
components, manipulation of which would be quite complicated in 5-dimensions. 
Since the vacuum equation is invariant under the interchange of Einstein 
and Ricci tensors, it would be invariant under the electrogravity duality 
transformation. Thus we will have to make use of the effective equation 
which is duality-variant to obtain dual solution. \\

 Let us consider the spherically symmetric metric in the Kaluza-Klein 
spacetime

\be
ds^2 = A^2 dt^2 - B^2 dr^2 -M^2 (d \theta^2 + \sin^2 \theta d \varphi^2) - C^2 
d \psi^2 
\ee

\noindent where $A,B,M,C$ are functions of $r$ only. We denote the 
coordinates as $x^0 = t, ..., x^4 = \psi$. For this metric the 
non-zero Ricci components in the basis frame are given by

\bq
R_{00} &=& \frac{1}{B^2} \bigg[-\frac{A^{\prime \prime}}{A} - \frac{A^{\prime}}{A} (\frac{2M^{\prime}}{M} - \frac{B^{\prime}}{B} + \frac{C^{\prime}}{C}) \bigg] \n \\
R_{11} &=& \frac{1}{B^2} \bigg[\frac{A^{\prime \prime}}{A} + 
\frac{C^{\prime \prime}}{C} 
+ \frac{2M^{\prime \prime}}{M} - \frac{B^{\prime}}{B} (\frac{2M^{\prime}}{M}
+ \frac{C^{\prime}}{C} + \frac{A^{\prime}}{A}) \bigg] \n \\
R_{22} &=& R_{33} = \frac{1}{B^2} \bigg[\frac{M^{\prime \prime}}{M} + 
\frac{M^{\prime }}{M} 
(\frac{A^{\prime}}{A} + \frac{C^{\prime}}{C} + \frac{M^{\prime}}{M}
- \frac{B^{\prime}}{B}) - \frac{B^2}{M^2} \bigg] \n \\
R_{44} &=& \frac{1}{B^2} \bigg[\frac{C^{\prime\prime}}{C} + 
\frac{C^{\prime}}{C} 
(\frac{2M^{\prime}}{M} + \frac{A^{\prime}}{A} - \frac{B^{\prime}}{B}) \bigg]
\eq 

 Note that in the Kaluza-Klein spacetime the vacuum solution unlike the 
Schwarzschild solution of GR is not unique. If all the metric 
coefficients are assumed to be given by powers of the same function;i.e. 
$A = f^a, B = f^b, C = f^c, M = rf^d, f = f(r)$. Then it turns out that for 
the vacuum solution it is sufficient to solve
\be
R^2_2 = 0, R^0_0 = R^1_1 = R^4_4
\ee
\noindent rather than all the Ricci components zero. The solution of this 
set would imply vacuum (all Ricci zero). This is the effective vacuum 
equation. This gives the vacuum solution,
\be
f = (1 - \frac{2m}{r})^{\frac{1}{2(d-b)}}, a+b+c=0, a(a+b)+d(2b-d)=0.
\ee
\noindent We thus have the vacuum solution [11] having two free dimensionless
parameters. \\
   
 Under the duality transformation Ricci goes to Einstein and hence the 
above effective equation (6) would go over to
\be
G^2_2 = 0, G^0_0 = G^1_1 = G^4_4 
\ee
\noindent which means
\be
R^0_0 = R^1_1 = R^4_4 = 0.
\ee

 This equation would under the same assumption admit the solution 
\be
f = (k -\frac{2m}{r})^{\frac{1}{2(d-b)}}
\ee
\noindent $a,b,c,d$ satisfying the same two relations as in (7). This is 
the solution obtained by Banerjee etal [10] for a global monopole charge in 
the Kaluza-Klein spacetime. The parameters, $m$ is the core mass and 
$k = 1 - \eta^2$ is due to the monopole field. 

  The metric for the dual solution would read as
\be
ds^2 = f^{2a} dt^2 - f^{2b} dr^2 - r^2f^{2d}(d \theta^2 + 
sin^2 \theta d\varphi^2) - f^{2c} d\psi^2
\ee
\n where
\be
f = (1 - \eta^2 - 2m/r)^{\frac{1}{2(d-b)}}
\ee
\noindent and $a,b,c,d$ are constrained by the same two relations as 
given in (7) above.

 This is the Banerjee etal solution [10] which is the  Kaluza-Klein 
analogue of the Barriola-Vilenkin solution [7] (which follows when $d = 0, 
a = 1$) for 4-dimensional global monopole charge.  The Gross-Perry vacuum 
solution follows when $\eta = 0$. It may be noted 
that unlike GR, neither the vacuum nor the dual-vacuum solutions are 
unique.

 It generates the stresses
\be
T^0_0 = T^1_1 = T^4_4 = \frac{1 - k}{M^2}
\ee
\noindent which disappear to go over to vacuum  when $k = 1$. The global 
monopole stresses take to this form asymptotically [10]. \\

 Let us show that the effective equation (6) and its dual (8) can be 
written in 
terms of energy and timelike and null convergence density. We define the 
energy density measured by a static observer by $\rho = T_{ab}u^au^b, 
u_au^a = 1$, and the timelike and null convergence density respectively 
by $\rho_t = 
(T_{ab} - 1/2 Tg_{ab})u^au^b$ and $\rho_n = T_{ab}v^av^b, v_av^a = 0$. 
The convergence density is responsible for focussing of timelike and null 
congruence as indicated in the Raychaudhuri equation [13]. \\

 Note that in 5-dimensions, 
\be
R_{ab} = -8\pi(T_{ab} - 1/3 Tg_{ab}).
\ee
 The dual equation is equivalent to $\rho_t = 0 = \rho_n$ where $\rho_n$ 
refers to non-angular coordinate geodesics. Note that $\rho_n = 0$ will 
imply $G_0^0 = G_1^1 = G_4^4$ and $\rho_t = 0$ will then give $G_2^2 = 
0$, yielding the dual equation (8). Apply the duality transformation, 
$R_{ab} \leftrightarrow G_{ab}$, to the above relation (14), then $\rho = 0 
=\rho_n$ gives the effective vacuum equation (6). 

 Similar to dual-vacuum solution obtained above (11), we can also obtain 
dual-flat solution. In 5-dimensions, flat space is effectively given by the 
equation \be
H_{[ab]} = 0 = \tilde E_{ab}, ~E_{ab} = \lambda(g^1_ag^1_b + g^4_ag^4_b)
\ee
\noindent where \\

\be
E^a_b = R^{a0}{}_{b0},
~~\tilde E^a_b = -E^a_b + P^a_b + P^0_0g^a_b - P^a_0g^0_b - P^0_bg^a_0
\ee

\noindent where

$$P^a_b = R^a_b - 1/4 Rg^a_b$$

The above relation for 
passive electric part follows from the general expression for the double 
dual of the Riemann tensor for $n\ge5$ dimensions [12]. Here the Riemann 
curvature is resolved relative to hypersurface orthogonal unit timelike 
vector. 

 From the above equation, we shall have $E_{22} = 0$ which would imply $A 
= const.$ (because $M$ cannot be constant) which would further imply 
$E_{ab} = 0$ leading to flat space because ${\tilde E}_{ab} = 0$.

 The dual-flat equation would follow by interchange of $E_{ab}$ and 
$\tilde E_{ab}$ and would read as
\be
H_{[ab]} = 0 = E_{ab}, ~\tilde E_{ab} = \lambda(g^1_ag^1_b + g^4_ag^4_b).
\ee
 
 \noindent Now $E_{ab} = 0$ gives $A = const. = 1$, $\tilde E_{22} = 0$ and 
$\tilde E^1_1 = \tilde E^4_4$ would give
$$
R^{12}{}_{12} + R^{14}{}_{14} + R^{24}{}_{24} = 0
$$
\noindent and
$$
R^{12}{}_{12} = R^{24}{}_{24}.
$$
  
 \n These two equations solve to give
\be
M^{\prime} = BC
\ee
\noindent and
\be
C^2 = k - \frac{2m}{M}.
\ee

 \n This is the dual-flat solution which is of course non-flat. Here the 
function $M$ remains undetermined.  It would give rise to the following 
stresses,
\be
T^0_0 = T^1_1 = T^4_4 = \frac{1 - k}{M^2}
\ee
\noindent which are exactly the same as the one given above. For $M = r$, 
the dual-flat solution would read as
\be
ds^2 = dt^2 - (k - 2m/r)^{-1} dr^2 - r^2 (d\theta^2 + sin^2\theta d\varphi^2) - 
(k - 2m/r) d\psi^2
\ee
\noindent which follows from the global monopole solution (11-12) when $a 
= d 
= 0, c = - b = 1$. Note that the dual-flat solution is the 4-dimentional 
Euclidean global monopole solution. Further when $m = 0$, the stress 
distribution would however remain undisturbed, we shall have zero-mass 
global monopole, which is the 5-dimentional analogue of the dual-flat 
solution [1,5]. Thus the dual-flat solution can have $m$ non-zero as well 
as zero.\\

 We have thus shown that the global monopole solution [10] is the 
dual-vacuum solution where the duality is defined by the transformation 
(3). That is interchange of active and passive electric parts which is 
equivalent to interchange of the 
Ricci and Einstein components (or $\rho\leftrightarrow\rho_t, 
\rho_n\rightarrow\rho_n$) in the effective vacuum equation (6). It is 
interesting that dual-flat solution in general gives rise to an Euclidean 
global monopole which has the zero mass limit analogus to the 
4-dimensional case. The duality transformation thus as in 4-dimensions 
puts on a global mnonopole charge on vacuum or flat spacetime. \\

Applications of the duality transformation have been considered
for non-empty spacetime [14], stringy black 
holes with dilaton field [15] and 2+1 gravity [9]. \\

{\bf Acknowledgement:} We thank Jose Senovilla for the expression (15). LKP 
and RS thank IUCAA for hospitality.


\newpage
    \begin{thebibliography}{99}

\bibitem{} N. Dadhich (1999) Mod. Phys. Lett. {\bf A14}, 337.

\bibitem{} -------- (1999) Mod. Phys. Lett. {\bf A14}, 79.

\bibitem{} -------- (1999) Gen. Relativ. Grav. (Contribution to George 
Ellis Festschrift), Submitted.

\bibitem{} -------- in {\it Black holes, Gravitational Radiation and 
the Universe}, eds. B. R. Iyer and B. Bhawal (Kluwer, 1999), p.171.

\bibitem{} N. Dadhich, Dual spacetimes, Mach's principle and 
topological defects, gr-qc/9902066.

\bibitem{} N. Dadhich, Empty space and its dual in general relativity, 
submitted.

\bibitem{} M. Barriola and A. Vilenkin (1989) Phys. Rev. Lett. {\bf 63},341.

\bibitem{} M. Nouri-Zonoz, N. Dadhich and D. Lynden-Bell (1999) 
 Class. Quantum Grav. {\bf16}, 1021.

\bibitem{} S. Bose, N. Dadhich and S. Kar, New classes of black holes in 
2+1 gravity, submitted.

\bibitem{} A. Banerjee, S. Chatterjee and A. A. Sen (1996) Class. Quantum 
Grav. {\bf 13}, 3141.

\bibitem{} J. D. Gross and M. J. Perry (1983) Nucl. Phys. {\bf B226}, 29.


\bibitem{} J. M. M. Senovilla (1999) Super energy tensors, to be submitted.

\bibitem{} A. K. Raychaudhury (1955) Phys. Rev. {\bf90}, 1123.

\bibitem{} N. Dadhich, L.K. Patel and R. Tikekar (1998) Class. Quanum
Grav. {\bf 15}, L27.

\bibitem{} S. Bose and N. Dadhich, Electrogravity-duality and global 
monopoles in scalar-tensor gravity, submitted.

   
\end{thebibliography}
\end{document}